\documentstyle[12pt]{article}
\newcommand{\sbody}[2]{{\textstyle\frac{#1}{#2}}}

\newcommand{\be}{\begin{equation}}
\newcommand{\ee}{\end{equation}\noindent}
\newcommand{\bear}{\begin{eqnarray}}
\newcommand{\ear}{\end{eqnarray}\noindent}
\newcommand{\no}{\noindent}
\date{}
\renewcommand{\theequation}{\arabic{section}.\arabic{equation}}

\newcommand{\Det}{{\rm Det}}

\newcommand{\slD}{\raise.15ex\hbox{$/$}\kern-.57em\hbox{$D$}}
\newcommand{\slpartial}{\raise.15ex\hbox{$/$}\kern-.57em\hbox{$\partial$}}
\newcommand{\slG}{{{\dot G}\!\!\!\! \raise.15ex\hbox {/}}}

\def\non{\nonumber}
\def\beqn*{\begin{eqnarray*}}
\def\eqn*{\end{eqnarray*}}

\def\square{\kern1pt\vbox{\hrule height 1.2pt\hbox{\vrule width 1.2pt
   \hskip 3pt\vbox{\vskip 6pt}\hskip 3pt\vrule width 0.6pt}
   \hrule height 0.6pt}\kern1pt}
\def\slash#1{#1\!\!\!\raise.15ex\hbox {/}}

\def\dps{\displaystyle}

\def\half{{1\over 2}}

\def\fourth{{1\over4}}

\def\e{\mbox{e}}

\def\kinb{{1\over 4}\dot x^2}

\def\4piTD{{(4\pi T)}^{-{D\over 2}}}
\def\4piT4{{(4\pi T)}^{-2}}

\def\Tintm4{{\dps\int_{0}^{\infty}}{dT\over T}\,e^{-m^2T}
    {(4\pi T)}^{-2}}
\def\Tintm{{\dps\int_{0}^{\infty}}{dT\over T}\,e^{-m^2T}}
\def\Tint{{\dps\int_{0}^{\infty}}{dT\over T}}

\def\Dx{\dps\int{\cal D}x}

\def\Dpsi{\dps\int{\cal D}\psi}
\def\Tr{{\rm Tr}\,}
\def\tr{{\rm tr}\,}

\begin{document}
\hskip12cm LAPTH-686/98\\
\vskip1pt
\begin{center}
\vfill
\large\bf{A New Approach to}\\
\large\bf{Axial Vector Model Calculations}
\end{center}
\vfill
\begin{center}
D.G.C. McKeon$^{(1)}$\\
Department of Applied Mathematics\\
University of Western Ontario\\
London CANADA\\
N6A 5B7\\
\vskip.5cm
Christian Schubert$^{(2)}$\\
Laboratoire d'Annecy-le-Vieux
de Physique de Particules\\
Chemin de Bellevue\\
BP 110\\
F-74941 Annecy-le-Vieux CEDEX\\
FRANCE
\end{center}
\vskip.9cm

{\large\bf Abstract}

\begin{quotation}

\noindent
We consider the one-loop effective action due to a spinor loop
coupled to an abelian vector and axial vector field background.
After rewriting this effective action in terms of an auxiliary
non-abelian gauge connection, we use the De Witt expansion
to analyze both its anomalous and non-anomalous content.  
The same transformation allows us to obtain a novel worldline
path integral representation for this effective action which
avoids the usual separation into the real and imaginary
parts of the Euclidean effective action, as well as the
introduction of auxiliary dimensions.
\end{quotation}
\noindent
11.15Bt

\vfill
email: $^{(1)}$TMLEAFS@APMATHS.UWO.CA/$^{(2)}$SCHUBERT@LAPP.IN2P3.FR\\

\eject
\pagestyle{plain}
 \setcounter{page}{1}

\section{Introduction}
\renewcommand{\theequation}{1.\arabic{equation}}
\setcounter{equation}{0}
\vskip-3pt

It is well known that if a spinor field $\psi$ is coupled to
background fields $A_\mu$ and $A_{5\mu}$ so that

\be
{\cal{L}} = \overline{\psi} (\slash p + \slash A +\gamma_5 
\not\!\!A_5 
-
m)\psi \quad\quad (p = -
i\partial)
\label{calL}
\ee
\no
then the axial current $j_\mu^5 = \overline{\psi} \gamma_\mu\gamma_5 
\psi$ has an anomalous
divergence \cite{schwinger51}.
One-loop processes can be analyzed in terms of the 
functional determinant

\be
i W_{eff}^{(1)} = \ln \Det (\slash p + \slash A +\gamma_5\not\!\!A_5 - 
m). 
\label{WMink}
\ee
\no
In Euclidean space, the corresponding
operator $H = \slash p + \slash A +
\gamma_5\not\!\!A_5  - im$ is not
Hermitian, and the anomaly can be attributed to the phase of the 
functional determinant appearing
in (\ref{WMink}) \cite{alvwit,mckshe}.  
In this paper, we avoid the separation of the effective action
into its real and imaginary part. Instead, we rewrite it
in terms of the effective action for a scalar loop in a certain
non-abelian background, and
use the De Witt
expansion \cite{dewitt} to discuss its anomalous and
non-anomalous content. (We could also work directly in Minkowski
space.)

The same transformation allows us to derive a novel 
worldline path integral
representation for the effective action.
While various such representations have been known and used
for decades in the case of pure vector amplitudes
\cite{feynman50,fradkin,casalbuoni}, generalizations
to amplitudes involving axial vectors and/or
pseudoscalars have been constructed only quite recently
\cite{mnss1,mnss2,dhogag,haasch}. Those were motivated by the
discovery of an efficient way of evaluating this type
of path integral \cite{strassler} 
which allows one, in particular, to recover many of the
calculational improvements which have been achieved by
representing field theory amplitudes as the infinite string tension
limits of appropriate string amplitudes \cite{berkos}
(see also \cite{mckeon} for the closely related ``Quantum Mechanical
Path Integral Formalism'').
However the previous approaches to axial vector models 
implied a separate treatment of the parity - even and - odd
parts of the effective action
\footnote{In \cite{dhogag} a way was found to combine both
parts into a single expression, however at the cost of 
introducing an
additional parameter integration.}
, and, in the latter case, the use of insertion
operators in analogy to superstring theory.
Here we will show that, for the special case where the background
consists
only of
a vector and axial vector field, 
there is a much simpler solution for this problem
which allows one to treat both parts of the
effective action on the same footing. 

\section{The One-Loop Euclidean Effective Action}
\renewcommand{\theequation}{2.\arabic{equation}}
\setcounter{equation}{0}

It is easily established that
\footnote{
In Euclidean space we use
$
\gamma_{Ej}\equiv i\gamma_j,
\gamma_{E4}\equiv \gamma_0,
\gamma_{E5}\equiv\gamma_5,
$
so that
$
\lbrace \gamma_{Ea},\gamma_{Eb}\rbrace =2\delta_{ab},
a,b=\mu,5
$. The subscript ``E''
is omitted in the following.
The Euclidean $\varepsilon$ - tensor is defined by
$\varepsilon_{1234} = 1$.
}

\be
(\slash p + \slash A +\gamma_5 \not\!\!A_5 )^2 
= \left( p_\mu +A_\mu -\gamma_5\sigma_{\mu\nu} A_5^\nu 
\right)^2 +
(D-2) A_5^2 +
iA_{5,\nu}^\nu \gamma_5 - {i\over 2}
\sigma_{\mu\nu} \left(\partial^\mu 
A^\nu - \partial^\nu
A^\mu \right)
\label{rewrite}
\ee
\no
($\sigma_{\mu\nu} \equiv \frac{1}{2} \left[ 
\gamma_\mu, \gamma_\nu
\right]$).  
Here we used the Dirac algebra in $D=4$ dimensions, dimensionally
continued with an anticommuting $\gamma_5$.
We can write (\ref{rewrite}) in the form

\be
(\slash p + \slash A +\gamma_5 \not\!\!A_5)^2 =
- (\partial_\mu 
+ i{\cal A}_\mu)^2 + a
\label{introAcal}
\ee\no
where

\bear
{\cal A}_{\mu} &\equiv& A_{\mu} -\gamma_5\sigma_{\mu\nu}A_5^{\nu}
\label{defcalA}\\
a &\equiv& -{i\over 2}
\sigma_{\mu\nu}
\Bigl(
\partial_{\mu}A_{\nu}-\partial_{\nu}A_{\mu}
\Bigr)
+i\gamma_5A_{5,\mu}^{\mu} + (D-2) A_5^2 
\label{defV}
\ear\no
Using the usual argument that

\be
{\rm Det}
\Bigl[(\not\!p + \not\!\!A +\gamma_5 \not\!\!A_5) - im\Bigr] = 
{\rm Det}
\Bigl[(\not\!p + \not\!\!A +\gamma_5 \not\!\!A_5) + im\Bigr] = 
\Det^{1/2}
\Bigl[(\not\!p + 
\not\!\!A +\gamma_5 \not\!\!A_5)^2 + m^2\Bigr]
\label{gamma5trick}
\ee\no
we can then write
the Euclidean effective action in the following form,

\be
W_{\rm{eff}}^{(1)} = -\half\, \Tr\, \int_0^\infty \, 
\frac{dT}{T} \, \exp 
\Bigl\lbrace
- T \left[-
(\partial_\mu + i{\cal A}_\mu)^2 + a + m^2 \right]
\Bigr\rbrace
\label{Weucl}
\ee\no
Up to the global sign, this is 
formally identical with the effective action for
a scalar loop in a background containing a 
non-abelian gauge field ${\cal A}$ and a potential $a$.
The operator in (\ref{Weucl}) 
is such that $W_{\rm{eff}}^{(1)} [{\cal A}_\mu, a]$ 
is invariant under the
transformations \cite{thooft}

\bear
{\cal A}_\mu &\rightarrow& 
{\cal A}_\mu + \partial_\mu \Lambda + 
i
[{\cal A}_\mu, 
\Lambda]\non\\
a &\rightarrow& a + i [a, \Lambda] 
\non\\
\label{gaugetrafo}
\ear\no
where the gauge parameter $\Lambda$ has values in the Clifford algebra.
However it should be noted that those transformations are
not identical with the transformations

\bear
A_\mu &\rightarrow& A_\mu + \partial_\mu \theta
\label{trafoA}\\
A_{5\mu} &\rightarrow& A_{5\mu} + \partial_\mu \theta_5
\label{trafoA5}
\ear\no
(unless $A_{5\mu} = 0$) 
as would be expected naively from (\ref{calL})
in the case where
$m = 0$.
To see the reason for this discrepancy, we examine carefully the
steps used in deriving (2.1).  This involves making the replacements

\begin{eqnarray}
(p \cdot \gamma + m \cdot \gamma)^2 = p^2 & + & \frac{1}{2} \left[ p_\mu
(\gamma_\mu m \cdot \gamma) + (\gamma_\mu m \cdot \gamma)p_\mu
\right] - \frac{i}{2} (\gamma_\mu m \cdot \gamma)_{, \mu} \nonumber \\
& + & \frac{1}{2}\left[ p_\mu (m \cdot \gamma \gamma_\mu) + (m \cdot
\gamma \gamma_\mu) p_\mu \right] + \frac{i}{2} (m \cdot \gamma
\gamma_\mu)_{, \mu} + (m \cdot \gamma)^2
\end{eqnarray}
and subsequently obtaining
\begin{eqnarray}
& = & (p_\mu + \frac{1}{2} \gamma_\mu m \cdot \gamma + \frac{1}{2} m
\cdot \gamma \gamma_\mu)^2 \nonumber \\
&& -  \frac{1}{4} (\gamma_\mu m \cdot \gamma + m \cdot \gamma
\gamma_\mu)^2 - 
\frac{i}{2} (\gamma_\mu m \cdot \gamma - m \cdot \gamma
\gamma_\mu)_{, \mu} + (m \cdot \gamma )^2
\end{eqnarray}
On the left side of eq. (2.10), the shift $p \rightarrow p + \partial
\Lambda$ can be compensated for by the replacement of $m$ by $m -
\partial \Lambda$.  This is not true on the right side of eq. (2.10)
unless $(\gamma_\mu \partial \Lambda \cdot \gamma - \partial \Lambda
\cdot \gamma \gamma_\mu)_{, \mu} = 0$ which holds when $\partial_\mu
\Lambda$ is identified with $\partial_\mu \omega$ where $\omega$ is a 
scalar
(which is what occurs if $m$ is the vector field $A_\mu$), but does
not hold when $\partial_\mu \Lambda$ is identified with $\partial_\mu
\omega \gamma_5$ (which is what occurs if $m$ is the axial vector
field $A_{5 \mu}$).  Consequently (2.7) and (2.9) are not equivalent.

\no
The De Witt expansion \cite{dewitt}

\be
\Tr\, e^{-H T} = \frac{1}{(4\pi T)^2} \, \sum_{n=0}^{\infty} \int dx \, 
a_n 
(x,x)T^n
\label{DeWitt}
\ee\no
can now be applied to the effective action in 
(\ref{Weucl}), and the known results \cite{blds,vandeven,fhss}
for the
non-abelian heat-kernel coefficients used.  The 
first few coefficients appearing in (\ref{DeWitt}) 
are given by 

\bear
a_1 &=& -\tr\left[a\right]\\
a_2 &=& \tr\left[- \sbody{1}{12} \, {\cal F}_{\mu\nu} {\cal F}^{\mu\nu} + 
\sbody12 \,
a^2\right]\label{a2}\\
a_3 &=&  -
\sbody{1}{12} \tr \left[2a^3 + S_\mu S^\mu  - a 
{\cal F}_{\mu\nu} {\cal F}^{\mu\nu}
-{4\over 15}i
{\cal F}_{\kappa\lambda}{\cal F}_{\lambda\mu}{\cal F}_{\mu\kappa}
+{1\over 10}
{\cal F}_{\kappa\lambda\mu}{\cal F}_{\kappa\mu\lambda}
\right]
\label{a3}
\ear\no
(${\cal F}_{\mu\nu} \equiv \partial_\mu{\cal A}_\nu - \partial_\nu 
{\cal A}_\mu +
i[{\cal A}_\mu, {\cal A}_\nu], \;\; S_\mu \equiv [\partial_\mu + 
i{\cal A}_\mu,
a],\;\;{\cal F}_{\alpha\beta\gamma} \equiv [\partial_\alpha + 
i{\cal A}_\alpha,
{\cal F}_{\beta\gamma}]$). 

\no
Performing the Dirac traces one obtains

\bear
a_1 &=& -4(D-2)A_5^2 \\
a_2 &=& \sbody{2}{3} \, F_{\mu\nu}F_{\mu\nu} 
+ \sbody43 \left[
(\partial_\mu A_{5\nu}) (\partial^\mu A_5^\nu) - (\partial \cdot
A_5)^2\right]\non\\
&=& 
\sbody{2}{3} \, 
\Bigl(
F_{\mu\nu}F_{\mu\nu}
+
F_{\mu\nu}^5
F_{\mu\nu}^5
\Bigr)
+
\quad
{\rm total}\quad{\rm derivative}\quad{\rm terms}
\label{computea2}
\ear\no
($F_{\mu\nu} \equiv \partial_\mu A_\nu - \partial_\nu A_\mu ; \;\; 
F_{\mu\nu}^5\equiv
\partial_\mu A_{5\nu} - \partial_\nu A_{5\mu}$)
\footnote{$a_1$ and $a_2$ have already been calculated in 
the context of quantum gravity with torsion
\cite{belsha}
(we thank I.L. Shapiro for this information).}.
Thus the vector field strength tensor appears automatically, 
but the
axial vector field strength tensor only after the addition of  
suitable total derivative terms to the effective Lagrangian.
Since the ultraviolet ($T = 0$) 
divergence is controlled by
$a_2$, we see from (\ref{a2}) that the divergent part of the effective 
action can be removed by counter
terms which are consistent with invariance under the transformations 
(\ref{trafoA}),(\ref{trafoA5}).

\no
Similarly, the contribution to $a_3$ from the three-point 
function $<AAA_5>$ is
given by
\footnote{The identification of this result with the
result of the heat kernel expansion in
standard field theory requires the use of the identity
$\varepsilon_{\alpha\beta\gamma\delta}
\Bigl(
F_{\alpha\beta,\lambda}F_{\gamma\lambda}V_{\delta}
+ F_{\alpha\beta}F_{\gamma\lambda,\lambda}V_{\delta}
+F_{\alpha\beta}F_{\lambda\gamma,\delta}V_{\lambda}
\Bigr) = 0$
, which holds true for an arbitrary field strength
tensor $F$ and vector $V$.}


\be
a_3^{\rm AAA_5}
=
i\epsilon^{\alpha\beta\gamma\delta}
\biggl\lbrace
F_{\alpha\beta}
\Bigl[
\half
F_{\gamma\delta} A_{5,\lambda}^{\lambda}
+{2\over 3}
F_{\lambda\gamma}
A_{5\delta ,\lambda}
\Bigr]
-{2\over 3}
F_{\alpha\beta ,\lambda}F_{\delta\lambda}A_5^{\gamma}
\biggr\rbrace
\label{a3VVA}
\ee\no
We see from (\ref{a3VVA}) that invariance under the transformation 
of (\ref{trafoA}) is 
retained in the $<AAA_5>$
process while that of (\ref{trafoA5}) is lost.  
Furthermore, there is no 
divergence in this anomalous 
three-point process at one-loop order, which is consistent with 
\cite{mann}. In the abelian case considered here all higher
$a_n$'s must be non-anomalous (for 
a discussion of the non-abelian effective
action see \cite{cogiso} and refs. therein).

Alternatively, eqs.(\ref{introAcal}),(\ref{gamma5trick})
could also be used for writing down a set of second-order
Feynman rules generalizing the ones of \cite{morgan},
and thus for scattering amplitude calculations.

We also note that by (2.2) and (2.15) the following
transformation can be made in the full generating functional for
a $U(1)$ axial gauge theory,

\begin{eqnarray}
&&\int d A_\mu^5 d \psi^+ d \psi 
\exp 
\biggl\lbrace -
\int dx \left( \frac{1}{4}
F_{\mu\nu}^5 F_{\mu\nu}^5 + \psi^+ (\slash p + \gamma_5 \not\!\!A_5)
\psi \right) 
\biggr\rbrace\non\\
&&
= \int d A_\mu^5 d \chi 
\exp 
\biggl\lbrace
-
\int dx \left[ \frac{3}{8} \tr
\left( - \frac{1}{12} {\cal
F}_{\mu\nu} {\cal F}_{\mu\nu} + \frac{1}{2} a^2 \right) + \chi^+ \left(
(p_\mu + {\cal A}_\mu)^2 + a \right) \chi \right]
\biggr\rbrace
\label{U1trafo}
\end{eqnarray}\no
($\chi^+ = \chi^T$).
The action now has the form of a ${\rm Spin}(4)$ gauge theory with a real
fermionic scalar field $\chi$.  

\section{Worldline Path Integral Representation of the Effective
Action}
\renewcommand{\theequation}{3.\arabic{equation}}
\setcounter{equation}{0}

In this section we will
transform eq.(\ref{Weucl}) into a first-quantized worldline path 
integral representation for the effective action. The standard
procedure for this transformation is the coherent state method
\cite{ohnkas}. 
It was applied to the present problem already by
D' Hoker and Gagn\'e \cite{dhogag}, 
albeit in a formalism based on
six - dimensional Dirac matrices, and with a separate treatment for
the real and the imaginary part of the effective action.

Note that the representation (\ref{Weucl}) yields the complete
effective action, real and imaginary part. The price which we have to
pay for this property is the non-hermiticity of the
kinetic operator in the exponent. However we still have positivity
of this operator for sufficiently weak background fields, which
is sufficient for perturbative purposes.
We refer the reader to \cite{dhogag} for a detailed account of the
coherent state method, and just present the final result of the
transformation,

\bear
W_{\rm{eff}}^{(1)} 
&=& -2\, \int_0^\infty \, 
\frac{dT}{T} 
\e^{-m^2T}
\Dx
\Dpsi
\, \e^
{
-\int_0^Td\tau\,
L(\tau)
}\non\\
L(\tau) &=&
\kinb
+\half\psi_{\mu}\dot\psi^{\mu}
+i\dot x^{\mu}A_{\mu}
-i\psi^{\mu}F_{\mu\nu}\psi^{\nu}
-2i\hat\gamma_5\dot x^{\mu}\psi_{\mu}\psi_{\nu}A_5^{\nu}
+i\hat\gamma_5\partial_{\mu}A^{\mu}_5
+(D-2)A_5^2
\non\\
\label{vapi}
\ear\no
Here $\int {\cal D}x$ 
denotes the coordinate path integral over the space
of all closed loops with fixed proper-time length $T$, and
$\int{\cal D}\psi$ 
a path integral over Grassmann - valued functions.
The periodicity properties of 
$\int{\cal D}\psi$ 
are determined by the
operator $\hat\gamma_5$; after expansion of the interaction
exponential a given term in the integrand will have to be
evaluated using antiperiodic (periodic) boundary conditions
on $\psi$, $\psi(T) = - (+)\,\psi(0)$, 
if it contains $\hat\gamma_5$
at an even (odd) power. After the boundary 
conditions are determined
$\hat\gamma_5$ can be replaced by unity.

The perturbative evaluation of the double path integral
can then be done as usual 
(see, e.g., \cite{schubert}) using worldline
Green's functions adapted to the periodicity 
conditions. For the coordinate path
integral one must first remove the zero mode contained
in the path integral, which may be done by fixing the
average position 
$x^{\mu}_0\equiv {1\over T}\int_0^Td\tau x^{\mu}(\tau)$
of the loop. The reduced path integral over the 
variable $y^{\mu}\equiv x^{\mu}-x^{\mu}_0$ is then
evaluated with a correlator

\be
\langle y^{\mu}(\tau_1)y^{\nu}(\tau_2)\rangle
=
-g^{\mu\nu}
\Bigl(
\mid \tau_1 - \tau_2\mid -
{{(\tau_1 - \tau_2)}^2\over T}
\Bigr)
\equiv
-g^{\mu\nu}
G_B(\tau_1,\tau_2)
\label{ycorr}
\ee\no
(other choices are also possible \cite{mckeon}).
With our conventions, the free Gaussian coordinate path
integral is, in dimensional regularization, equal to

\be
\Dx\,\,\e^{-\int_0^Td\tau\,\kinb}
=
{(4\pi T)}^{-{D\over 2}}\int dx_0
\label{freeypi}
\ee
\no
For the Grassmann path integral one has to proceed differently
depending on the boundary conditions.
In the antiperiodic case (``A'') there is no zero mode, so that
$\int_A {\cal D}\psi$ 
can be executed straightforwardly with a worldline
correlator

\be
\langle \psi^{\mu}(\tau_1)\ \psi^{\nu} (\tau_2)\rangle
=  \, \half g^{\mu\nu} 
  {\rm sign}
(\tau_1 -\tau_2 )
\equiv
\half g^{\mu\nu}G_F(\tau_1,\tau_2)
\label{psicorr}
\ee\no
In the periodic case (``P'')
one has again a zero mode which must be separated
off, so that 

\bear
\int_P{\cal D}\psi &=& \int d\psi_0\int {\cal D}\xi \non\\
\psi^{\mu}(\tau)&=&\psi_0^{\mu} + \xi^{\mu}(\tau)\non\\
\int_0^Td\tau\,\xi(\tau) &=& 0\non\\
\label{splitgrass}
\ear\no
The zero mode integration produces an $\varepsilon$ - tensor via

\be
\int d^4\psi_0
\psi_0^{\mu}\psi^{\nu}_0\psi^{\kappa}_0\psi^{\lambda}_0 
=\varepsilon^{\mu\nu\kappa\lambda}
\label{zeromodeintegral}
\ee\no
and the $\xi$ - path integral can be performed using the correlator

\be
\langle\xi^{\mu}(\tau_1)\, \xi^{\nu}(\tau_2)\rangle
= g^{\mu\nu}
\Bigl(
\half {\rm sign}(\tau_1 -\tau_2 )
-
{\tau_1 -\tau_2 \over T}  
\Bigr)
=
\half
g^{\mu\nu}
\dot G_B(\tau_1,\tau_2)
\label{xicorr}
\ee\no
(a ``dot'' always refers to a derivative with respect
to the first variable).
The free Grassmann path integrals are normalized to unity
in the antiperiodic, and to ${1\over 4}$ in the periodic case.

We have verified for a number of cases that 
the path integral representation (\ref{vapi})
indeed reproduces
the correct field theory results,
for example the coefficient
$a_3^{AAA_5}$ of eq.(\ref{a3VVA}).
We remark that for the calculation of divergent
amplitudes 
it is essential to keep the coefficient of the
$A_5^2$ - term in the worldline Lagrangian
$D$ - dependent, if one wishes to use this formalism together
with dimensional regularization.
This can be seen already in the case of the (massive) 
$\langle A_5 A_5\rangle$ amplitude, where this term
produces a tadpole
contribution which is necessary to obtain
the same result as is reached in
the corresponding
field theory calculation (performed
with a naive anticommuting $\gamma_5$).

\section{Worldline Path Integral Calculation of the ABJ Anomaly}
\renewcommand{\theequation}{4.\arabic{equation}}
\setcounter{equation}{0}

Finally, we apply the 
vector - axial vector path integral 
to yet another recalculation of the
chiral anomaly, this time in momentum space.
The usefulness of first-quantized path integrals
for the calculation of anomalies was already
established in \cite{alvarezgaume,petervan}
(see also the $D=2$ calculation in \cite{mnss2}
which is more closely related to the following one).
Thus we would like to calculate the anomalous divergence
of the axial current in the $\langle AAA_5\rangle$
amplitude, which in field theory would be given in terms
of a sum of two triangle diagrams.
To extract this amplitude from the effective action, as usual
we must specialize the background fields to plane waves,

\bear
A^{\mu}(x)&=& \varepsilon_1^{\mu}\e^{ik_1\cdot x}
             +\varepsilon_2^{\mu}\e^{ik_2\cdot x}
\non\\
A_5^{\mu}(x)&=&\varepsilon_3^{\mu}\e^{ik_3\cdot x}
\label{decompAA5}
\ear
and then keep the part of the effective action which is linear
in all polarization vectors. In particular the 
$A_5^2$ -
piece in the worldline Lagrangian will not yet contribute at
this level.
Thus we find 
the following representation for the three-point function,

\bear
\langle A^{\mu}[k_1]A^{\nu}[k_2]A_5^{\rho}[k_3]\rangle
&=&
-2i
\Tintm
\Dx\Dpsi
\,\,\exp
\biggl\lbrace
-\int_0^Td\tau
\,
\Bigl(
\fourth
\dot x^2 + \half \psi\cdot\dot\psi
\Bigr)
\biggr\rbrace
\non\\
&&\times
\int_0^Td\tau_1
\Bigl(\dot x^{\mu}_1+2i\psi^{\mu}_1
k_1\cdot\psi_1
\Bigr)
\e^{ik_1\cdot x_1}
\int_0^Td\tau_2
\Bigl(\dot x_2^{\nu}+2i\psi^{\nu}_2
k_2\cdot\psi_2
\Bigr)
\e^{ik_2\cdot x_2}
\non\\
&&\times
\int_0^Td\tau_3
\Bigl(
ik_3^{\rho}
+2\psi^{\rho}_3\dot x_3\cdot\psi_3
\Bigr)
\e^{ik_3\cdot x_3}
\label{AAA5amplitude}
\ear
where the Grassmann path integral is periodic in this case.
It is important to note here the following two facts.
Firstly, this expression represents not a single triangle
diagram but the sum of the two. 
Secondly, this expression is already manifestly gauge invariant,
i.e. transversal in the vector current indices. If one multiplies
the right hand side by, say, $k_1^{\mu}$, then the 
photon vertex 
operator representing leg 1
becomes the integral of
a total derivative,
which vanishes
due to periodicity. This mechanism is, of course, well-known from
string theory. Nothing analogous holds true for the axial-vector
vertex operator.
Thus the structure of the path integral eq.(\ref{vapi})
already forces the divergence of the 
vector current to vanish, and we clearly have to look at the
axial vector current to find the anomalous divergence.
We are
interested in this divergence only, rather than in a calculation
of the complete amplitude,
so that we can simplify
by contracting eq.(\ref{AAA5amplitude}) with $k_3^{\rho}$.
Also we can put legs $1$ and $2$ on-shell,
$k_1^2=k_2^2=0$, and restrict ourselves to the massless
case.

Since for this amplitude the Grassmann path integral has periodic
boundary conditions, according to the above we have
to rewrite 
$\psi_i^{\alpha}(\tau) =\psi_{0i}^{\alpha} + \xi_i^{\alpha}(\tau)$,
and then to keep only those terms which contain four factors of
the zero mode piece $\psi_0$. Using 
eqs.(\ref{ycorr}),(\ref{zeromodeintegral}),(\ref{xicorr})
we obtain (deleting the
energy-momentum conservation factor)

\bear
k_3^{\rho}
\langle A^{\mu}A^{\nu}A_5^{\rho}\rangle
&=&
-2
\varepsilon^{\mu\nu\kappa\lambda}k_1^{\kappa}k_2^{\lambda}
\Tint
{(4\pi T)}^{-2}
\prod_{i=1}^3\int_0^Td\tau_i
\exp\Bigl\lbrack
(G_{B12}-G_{B13}-G_{B23})
k_1\cdot k_2
\Bigr\rbrack
\non\\
&&\times
\biggl\lbrace
\Bigl[
2+
(\dot G_{B12} +\dot G_{B23} +\dot G_{B31})
(\dot G_{B13}-\dot G_{B23})
\Bigr]
k_1\cdot k_2
-(\ddot G_{B13} +\ddot G_{B23})
\biggr\rbrace
\non\\
\label{divAAA5}
\ear
Here momentum conservation has been used to eliminate
$k_3$, and we abbreviate 
$G_{Bij}\equiv G_{B}(\tau_i,\tau_j)$ etc. 
We 
remove the second derivatives
$\ddot G_{B13}$ ($\ddot G_{B23}$) by a
partial integration in $\tau_1$ ($\tau_2$).
The expression in brackets then turns into

\be
k_1\cdot k_2
\,
\biggl\lbrace
2- (\dot G_{B12}+\dot G_{B23}+\dot G_{B31})^2
+\dot G_{B12}^2
-\dot G_{B13}^2
-\dot G_{B23}^2
\biggr\rbrace
=
4{k_1\cdot k_2\over T}(G_{B13}+G_{B23}-G_{B12})
\label{rewritebraces}
\ee
(Here the identities
$
\dot G_{Bij} + \dot G_{Bjk} + \dot G_{Bki}
=
-{\rm sign}(\tau_i -\tau_j)
{\rm sign}(\tau_j -\tau_k)
{\rm sign}(\tau_k -\tau_i)
$
and
$
\dot G_{Bij}^2 = 1-4{G_{Bij}\over T}
$
are useful).
But 
this is precisely the same expression which appears also in the
exponential factor in (\ref{divAAA5}). After a rescaling
to the unit circle, and performance of the trivial $T$ - integral,
we find therefore a complete cancellation between the Feynman
numerator and denominator polynomials
\footnote{This cancellation occurs even if one does not put
legs 1 and 2 on-shell.}
.
Thus without further integration we obtain
already the desired result for the anomalous divergence,

\be
k_3^{\rho}
\langle A^{\mu}A^{\nu}A_5^{\rho}\rangle
=
-{8\over {(4\pi)}^2}
\varepsilon^{\mu\nu\kappa\lambda}k_1^{\kappa}k_2^{\lambda}
\label{PCAC}
\ee

\section{Discussion}
\renewcommand{\theequation}{5.\arabic{equation}}
\setcounter{equation}{0}

We have shown that the effective action for a spinor loop
coupled to a vector and axial vector background can be
reexpressed in terms of an auxiliary
nonabelian gauge field and potential.
This has allowed us to use known results for the
DeWitt expansion, and also to
discuss
the chiral anomaly from a novel point of view. 
Indeed, by the form of 
eq. (\ref{rewrite}), it is evident that an
anomaly will be generated in the divergence of any symmetry 
corresponding to invariance under
a chiral gauge transformation in the original action.

Moreover, we have derived a worldline path integral representation
for this effective action which, while similar to previous
proposals, has some obvious advantages. 
In field theory terms it corresponds to a naive anticommuting
treatment of $\gamma_5$ in dimensional regularisation.
However in contrast to the NDR scheme in field theory
it fixes the ABJ anomaly in such a way that the anomalous
currents are confined to the axial vectors.

The method employed generalizes in an obvious way to the inclusion
of an additional scalar field, as well as to the non-abelian
case. This will be discussed in a separate publication,
where we will also present some
more extensive calculations.

It would be interesting to pursue this approach in conjunction with 
anomalies in higher dimensions \cite{frakep} and the 
diffeomorphism anomalies 
\cite{alvwit}.

\vskip12pt
\no
{\bf Acknowledgements}
We would like to acknowledge helpful conversations with D. Macdonald,
I. Sachs, and R. Stora, as well as computer support by
D. Fliegner.  
NSERC provided financial support.

\end{document}